\documentclass[pra, twocolumn]{revtex4}

\begin{document}

\newcommand{\ket}[1]{|{#1}\rangle}

\title{Fault-Tolerant Quantum Computation}
\author{Daniel Gottesman}
\email[E-mail: ]{dgottesman@perimeterinstitute.ca}
\affiliation{Perimeter Institute for Theoretical Physics, Waterloo,
ON N2L 2Y5, Canada}

\begin{abstract}
I give a brief overview of fault-tolerant quantum computation, with
an emphasis on recent work and open questions.
\end{abstract}

\maketitle

The world is a dangerous place, particularly if you are a qubit.
Organized vibrations (or just thermal phonons) are constantly trying
to shake you up for all you're worth. Gangs of unemployed photons
fly about, flipping over any qubit foolish enough to get in their
way. A moment's relaxation can send you spontaneously plummeting to
your ground state.  And as for having a friend keep an eye on you,
forget about it --- a qubit being watched is no better than a
conventional classical bit.

Indeed, it is unlikely that a single qubit can survive for long on
its own.  But by teaming up as a quantum error-correcting code
(QECC), groups of qubits can work together to fight off the dangers
of their environment.  Indeed, by acting in concert, the qubits of a
QECC can not only survive, but flourish, performing together complex
quantum computations without losing their integrity.  In order to do
so, however, all their interactions need to be carefully structured
according to the dictates of a fault-tolerant protocol.  If the
qubits break this code of behavior, they run the risk of, well,
breaking the code, exposing them once again to the chaotic
environment of the world outside the quantum computer.

The basic techniques of quantum error correction and fault-tolerant
quantum computation were first developed in 1995--6, in response to
the surge of interest in quantum computation that followed Shor's
1994 factoring algorithm.  Those theoretical results showed that
there was unlikely to be any barrier in principle to building large
quantum computers, although of course serious practical difficulties
remain even today.  After the breakthroughs of the mid-90s, the
field experienced a period of relative inactivity, but in the past
few years there has been a burst of new research aimed mostly at
bringing the theory closer in various ways to the experimental
reality.

Fault tolerance is not the only approach that has been discovered to
deal with errors, but it is the most general one.  A properly
designed fault-tolerant protocol can correct arbitrary small errors
(the precise meaning of a ``small error'' will be discussed in
section~\ref{sec:FTcond}).  Other sorts of error control techniques,
such as the self-correcting pulses now employed in a variety of
different quantum computer implementations, tend to be good for
certain kinds of errors, but will leave other residual errors
behind.  Given the extremely high accuracy required for a large
quantum computation, it is likely that fault-tolerant quantum
computation will be needed to correct these remaining errors in
order for the computation to succeed.  Thus, it seems likely that
future quantum computers will use a variety of specific control
techniques to eliminate the most prevalent types of errors and will,
on top of that, employ fault-tolerant protocols to eliminate the
remaining sources of noise.

\section{Quantum Error-Correcting Codes}
\label{sec:QECC}

The first element in a fault-tolerant protocol is the quantum
error-correcting code.  Space does not permit a full exposition here
of the principles of quantum error correction.  Instead, I will just
present one specific code, the $7$-qubit code, for which
fault-tolerant operations are particularly straightforward.

The $7$-qubit code encodes a single qubit as follows:
\begin{eqnarray}
\ket{\overline{0}} \!\! & = & \!\! \ket{0000000} + \ket{1111000} +
\ket{1100110} + \ket{1010101} \\ & & \ + \ket{0011110} +
\ket{0101101} + \ket{0110011} +
\ket{1001011} \nonumber \\
\ket{\overline{1}} \!\! & = & \!\! \ket{1111111} + \ket{0000111} +
\ket{0011001} + \ket{0101010} \\ & & \ + \ket{1100001} +
\ket{1010010} + \ket{1001100} + \ket{0110100} \nonumber
\end{eqnarray}
The encoded zero state (or {\em logical zero}) $\ket{\overline{0}}$
is the superposition of the even weight codewords from the classical
$7$-bit Hamming code, whereas the logical one $\ket{\overline{1}}$
is the superposition of the odd weight codewords from the Hamming
code.  The {\em weight} of a codeword is the number of $1$s in the
codeword.  The classical Hamming codes have the useful property that
to get from any codeword in the code to any other codeword, you must
flip at least $3$ bits.  Thus, if we are given a codeword with only
one flipped bit (or no flipped bits) then we can identify
unambiguously which bit was flipped and what the original codeword
was, and the Hamming code can correct one error on an arbitrary bit.

It is a bit less straightforward to see how the $7$-qubit code can
correct a single error on an arbitrary qubit.  First of all, note
that an arbitrary encoded state $\alpha \ket{\overline{0}} + \beta
\ket{\overline{1}}$ is a superposition of codewords from the
classical Hamming code.  Therefore, if the code experiences a bit
flip or $X$ error ($\ket{0} \rightarrow \ket{1}$, $\ket{1}
\rightarrow \ket{0}$) on a specific qubit, the state is a
superposition of Hamming codewords which have been flipped at the
corresponding bit location. By the error-correcting properties of
the Hamming code, it is therefore possible to make a measurement to
determine the location of the error.  However, it is critical that
the measurement we make does {\em not} tell us what the original
classical codeword was, as that would destroy the quantum
superposition that gives us an encoded qubit rather than an encoded
classical bit.  One way to do this measurement will be discussed in
section~\ref{sec:FTprotocols}.

As it happens, the code can be used to correct phase errors in
nearly the same way.  Because of the particular superpositions of
classical codewords chosen in the encoding, if we perform a Hadamard
transform ($\ket{0} \rightarrow \ket{0} + \ket{1}$, $\ket{1}
\rightarrow \ket{0} - \ket{1}$) on each of the physical qubits
comprising the code, we again get a superposition of codewords from
the classical Hamming code. The Hadamard transform will convert a
phase flip or $Z$ error ($\ket{0} \rightarrow \ket{0}$, $\ket{1}
\rightarrow - \ket{1}$) into an $X$ error, so by checking for errors
in the Hadamard-transformed basis, we can identify the location of
any single $Z$ error.

The phase correction procedure can be done independently of
measuring the location of an $X$ error, so the code can therefore
correct not just $X$ and $Z$ errors, but also $Y = iXZ$ errors, or
indeed one $X$ error and one $Z$ error on different qubits.  (The
global phase factor $i$ has no physical significance, but it makes
some of the mathematics nicer.)  In addition, in the case that there
is no error (the identity $I$), the state is kept safe as well, and
is not destroyed by our error correction procedure.  The four types
of errors $I$, $X$, $Y$, and $Z$, acting on the $n$ qubits of the
code, generate a group of possible errors, frequently known as the
Pauli group.  We can define the {\em weight} of an operation from
the Pauli group analogously to the weight of a classical bit string
as the number of the qubits on which the operation acts
non-trivially (i.e., the total number of $X$s, $Y$s, and $Z$s that
appear in the tensor product of the operator).  The $7$-qubit code
can thus correct an arbitrary weight $0$ or $1$ Pauli operator.

However, $X$, $Y$, and $Z$ are not the only possible things that can
go wrong on a single qubit.  The most general possible error would
be described by some unitary interaction between the faulty qubit
and some environment.  However, $I$, $X$, $Y$, and $Z$ form a basis
for the set of $2 \times 2$ matrices, so any unitary $U$ can be
written as
\begin{equation}
U = \alpha I \otimes A_I + \beta X \otimes A_X + \gamma Y \otimes
A_Y + \delta Z \otimes A_Z,
\end{equation}
with $A_I$, $A_X$, $A_Y$, and $A_Z$ operators acting on the
environment.  Suppose we were to forget about the possibility of
this general error, and instead just perform error correction as if
only Pauli errors could occur. Quantum mechanics is linear, and the
error correction procedure accurately records any Pauli errors, so
the complete state of the system will be entangled between the
encoded state, the environment, and the extra register (the {\em
ancilla}) which we are using to record information we have measured
about the error, and will be a superposition of cases where the
incorrect qubit has each possible Pauli error or no error at all,
and where the ancilla's state correctly records which error is
present.  But then when we measure the ancilla, we collapse not just
the ancilla itself, but also the data and the environment, into the
case where one particular Pauli error $I$, $X$, $Y$, or $Z$ has
occurred.  Our na\"ive belief that only a Pauli error was possible
has become a self-fulfilling prophecy.

Of course, this $7$-qubit code only encodes one qubit, but if we
wish to encode many qubits, we can simply use this code repeatedly,
expanding each logical qubit in our system into $7$ physical qubits.
Each set of $7$ qubits is called a {\em block} of the code, and this
system is capable of correcting one error in each block.  Because of
the linearity of quantum mechanics, entanglement between separate
blocks is protected from errors as well.  Of course, if we happen to
have two errors in a given block, that block may fail, and if we
attempt to decode the failed block, we will get the wrong quantum
state out.  However, when errors are somewhat rare, two errors close
together are even rarer, so a QECC can convert a small physical
error rate into an even smaller logical error rate.

\section{Fault-Tolerant Protocols}
\label{sec:FTprotocols}

A quantum error-correcting code by itself is only useful when we can
consider the quantum gates we perform to be effectively perfect.
That is unlikely to be a good approximation for all but the smallest
quantum computations, so we must supplement the QECC with a
fault-tolerant protocol.  A fault-tolerant protocol must contain
components to perform fault-tolerant error correction,
fault-tolerant state preparation, fault-tolerant measurement, and a
universal set of fault-tolerant gates.  Typically, completing this
set involves a rather large collection of different tricks, but most
of them are devoted to conquering the same problem: error
propagation. Usually, we assume that a single faulty gate can cause
errors only in the qubits involved in the gate, so a bad
single-qubit gate would cause errors in at most one qubit and a
two-qubit gate in at most two qubits. However, even a perfect
two-qubit gate can propagate a pre-existing error from one of the
qubits involved in the gate to the other one, so even a single
erroneous gate can thus indirectly cause errors in many qubits.

A QECC can only correct a limited number of errors per block (one in
the case of the $7$-qubit code described above), so we must be
particularly careful that errors do not spread within a block.  The
usual solution to this is to use {\em transversal} gates --- gates
which interact only the $i$th qubit of one block with the $i$th
qubit of another block.  With a system composed completely of
transversal gates, a single bad gate anywhere in the system can only
spread to produce one error per block of the code, which avoids
overwhelming the error tolerance of any single block.  Of course,
{\em two} bad gates could then cause a large number of blocks to
fail, so we must also perform error correction periodically to get
rid of the errors before they build up.

The $7$-qubit code described in section~\ref{sec:QECC} is
particularly convenient for fault-tolerance because a number of
gates can be performed transversally.  In particular, the logical
Hadamard transform can be performed by just performing the Hadamard
transform on each of the $7$ physical qubits in the code, and the
logical CNOT gate ($\ket{a}\ket{b} \rightarrow \ket{a} \ket{a \oplus
b}$, with $a$ and $b$ bits) can be performed transversally with CNOT
gates from each of the $7$ qubits of the control block to the
corresponding qubits in the target block.  The logical $\pi/4$ phase
rotation, $\ket{0} \rightarrow \ket{0}$, $\ket{1} \rightarrow i
\ket{1}$, can be done by performing the $-\pi/4$ phase rotation on
each physical qubit. Products of these gates are equally easy; for
instance, the $Z$ gate is the square of the $\pi/4$ gate and can be
performed simply by a $Z$ gate on each qubit, and similarly for $X$
and $Y$. The theory of fault tolerance makes frequent reference to
these gates and to the group (frequently called the {\em Clifford
group}) formed by all possible products of the CNOT, the Hadamard,
and the $\pi/4$ gate.

Unfortunately, the Clifford group is not universal for quantum
computation, and only the gates from the Clifford group can be
performed transversally on the $7$-qubit code.  Other transversal
combinations of gates would take a correct codeword to one with
errors in it. Adding even a single additional gate to the mix is
sufficient, however, to allow us to approximate any unitary
operation to arbitrary accuracy. A frequently used additional gate
is the $\pi/8$ gate, $\ket{0} \rightarrow \ket{0}$, $\ket{1}
\rightarrow \exp(i \pi/4) \ket{1}$.  This gate is usually performed
by first constructing a special ancilla state consisting of the
encoded state $(\ket{\overline{0}} + \exp(i \pi/4)
\ket{\overline{1}})/\sqrt{2}$, which in some sense encapsulates the
$\pi/8$ gate, and then performing a technique akin to quantum
teleportation to move the gate onto the data.  The teleportation
technique involves only Clifford group gates and measurement
(described below), so can be done transversally.  The ancilla
preparation requires some more complicated tricks, however, which I
will not describe here due to lack of space.

Fault-tolerant measurement is straightforward for the $7$-qubit
code.  Since the logical codewords are superpositions of classical
Hamming codewords, measuring each qubit gives us (in the absence of
errors) a random classical codeword, with an even-weight codeword
corresponding to a logical zero and an odd-weight codeword
corresponding to a logical one.  If there is an error, we may
instead get a classical codeword with a bit flip in it, but that can
be corrected easily using classical error correction.  This is a
destructive measurement procedure, but if we wish to perform a
non-demolition measurement, we can prepare an encoded ancilla in the
state $\ket{\overline{0}}$, perform a logical CNOT from the data
block to be measured to the ancilla, and then destructively measure
the ancilla.

Fault-tolerant state preparation is a bit more complicated, and a
variety of ways have been suggested to do it.  Fault-tolerantly
encoding an unknown quantum state is not possible, since a single
error at the very start of our encoding procedure will ruin the
state regardless of how cleverly we design the rest of the
procedure.  Instead, we typically perform fault-tolerant quantum
computations by first creating a number of encoded
$\ket{\overline{0}}$ states and using classical control and
fault-tolerant gates to produce the correct initial state for our
algorithm.  We thus need only consider how to fault-tolerantly
produce $\ket{\overline{0}}$ states.

We can start with a non-fault-tolerant encoding circuit, for
instance, but even a single error in the procedure could produce
another state, such as the $\ket{\overline{1}}$.  One way to avoid
this is to produce two $\ket{\overline{0}}$ states and check them
against each other using the non-demolition measurement procedure
described above.  If one of the states is an encoded
$\ket{\overline{1}}$ while the other is a $\ket{\overline{0}}$, the
measurement will identify that there is a problem (although it does
not tell us which one was wrong), and we discard both and try again.
Of course, if {\em both} states have errors in the preparation
procedure, we may be fooled into accepting a bad state, but that
requires two separate errors, which could ruin the state anyway.

Finally, we can perform fault-tolerant error correction by combining
the fault-tolerant preparation, measurement, and Clifford group
constructions from above.  Using fault-tolerant preparation and
Clifford group gates we can make reliable $\ket{\overline{0}}$ and
$\ket{\overline{0}} + \ket{\overline{1}}$ ancilla states.  Note that
a logical CNOT with the data block as a control block and a
$\ket{\overline{0}} + \ket{\overline{1}}$ ancilla as the target
would not have any effect at all in the absence of any errors.
However, if there are $X$ errors in the data, this operation will
cause them to propagate to the corresponding locations in the
ancilla block.  Then a transversal measurement of the ancilla will
give us a random classical Hamming codeword with bit flip errors in
the locations corresponding to the $X$ errors in the data, and if
there is only one such error, we can then identify its location
using classical error correction.  Note that because we used an
ancilla in the state $\ket{\overline{0}} + \ket{\overline{1}}$, the
measurement gives us no information about the encoded state of the
data, only the errors on the physical qubits making up the data
block.

Similarly, performing a logical CNOT from a $\ket{\overline{0}}$
ancilla to a data block copies the phase errors without disturbing
the logical data qubit.  Then measuring the ancilla in the Hadamard
basis again gives a random Hamming codeword with errors in the
places corresponding to the locations of $Z$ errors in the data
block. Of course, while errors are propagating from the data block
to the ancilla block, they are also propagating the other way, from
the ancilla blocks into the data block.  This procedure can never
make the data more reliable than the ancillas, but the ancillas we
use are always built from scratch, so their error rates are never
too high, and fault-tolerant error correction will prevent the
frequency of errors in the data blocks from building up to a level
where they are likely to cause our computation to fail.

We can understand fault-tolerant error correction from a
thermodynamic point of view: Errors introduce entropy, heating up
the state of our computer.  We introduce cool ancilla states, and
error correction pumps heat from the data into the ancillas, acting
like a refrigerator.  We can never cool the data to below the
temperature of the ancillas, but we can prevent it from heating up
to arbitrarily high temperatures.

\section{The Threshold for Fault Tolerance}

Of course, the $7$-qubit code has a definite limit to its
usefulness.  If two errors occur in a block (either directly or by
propagating from another block) before we have the opportunity to do
error correction, the block will fail, potentially introducing an
error in the state we were trying to protect.  If the probability
for a single error in a single physical gate (or a time step without
a gate) is $p$, then the probability of having two errors in two
particular physical gates is $p^2$.  However, by encoding the state
and by using a fault-tolerant protocol, we have introduced many
extra places something could go wrong, so the total probability of
having two errors accumulate in a block despite our attempts at
error correction becomes something like $Cp^2$ per logical gate,
where $C$ represents roughly the number of pairs of locations (gates
or time steps) where two physical errors can cause a logical error
in the course of performing the gate. $C$ will depend on the code
and the fault-tolerant circuitry we use, but if we perform error
correction at regular intervals, $C$ will not depend on the length
of the overall computation.

Now, replacing a physical error rate of $p$ per gate with a logical
error rate of $Cp^2$ per gate is an improvement when $p < 1/C$
(although it is actually worse when $p > 1/C$; in that case, the
extra qubits and gates we introduce cause extra errors faster than
we can correct them).  Before, we could do a computation of length
about $1/p$ before we would expect to see an error, and now we can
last for a time about $1/(Cp^2)$.  However, we would like to do
better still.

One way to achieve this is to use {\em concatenated codes}.  We can
encode each logical qubit in $7$ physical qubits, as before, but
then we can encode each of those $7$ qubits again using the same
code (or a different one, if we so desire).  Now the logical qubit
is encoded by $49$ physical qubits, but the error rate has dropped
again, to $C (Cp^2)^2$, or $C^3 p^4$.  If necessary, we can encode a
third or fourth time, giving a logical error rate $O(p^8)$ or
$O(p^{16})$. After $k$ levels of concatenation, the effective
logical error rate is
\begin{equation}
p_k = p_T (p/p_T)^{2^k},
\end{equation}
where $p_T = 1/C$ is called the {\em threshold}.  When $p<p_T$, the
logical error rate can be made arbitrarily small.  To achieve an
error rate $\epsilon$, we need $\log \log 1/\epsilon$ levels of
concatenation; of course, the number of qubits in a concatenated
block is exponential in the number of levels, but that still means
that we need only ${\rm poly}(\log 1/\epsilon)$ extra qubits to
achieve error rate $\epsilon$.

This result is known as the threshold theorem.  A complete rigorous
proof is more difficult than the above argument, as it must define,
for instance, the meaning of error rate in a code block which is
itself part of a larger concatenated code block, but the conclusion
is the same: There is a threshold error rate such that, if the
physical error rate per gate and per time step is below the
threshold value, then arbitrarily long universal quantum
computations are possible with only polylogarithmic overhead.  A
word of caution is necessary, however, regarding the overhead:
Estimates for plausible error rates and computation lengths
frequently require overhead of 1000:1 or more using concatenation.
The advantage of the threshold result is in the scaling --- much
longer computations require only slightly more overhead.

\section{The Value of the Threshold}
\label{sec:FTcond}

As you may imagine, a considerable amount of research has been
devoted to determining the numerical value of the threshold, as that
sets a target value of accuracy that experimentalists will try to
achieve. However, citing a single number as the threshold value is a
bit deceptive, as there are a large number of variables involved.
Also, it is important to bear in mind that thresholds are derived
for specific QECCs and fault-tolerant protocols.  Future
developments in these areas might increase our estimates of the
threshold, allowing fault-tolerant quantum computation with noisier
devices.

Threshold calculations frequently make a number of assumptions about
the properties of the quantum computer; while any given
implementation may satisfy some of these, I don't know of any that
satisfies all of them.  Below, I list some common assumptions worth
discussing further.  These are not the only assumptions necessary to
prove the threshold theorem.  Indeed, they are interesting to
discuss precisely because none of them is completely necessary.
Because we may not be able to satisfy all of these assumptions at
once, it is important to study the tradeoffs between them.
\begin{enumerate}
\item {\it Multiple-qubit gates can interact any pair of qubits in the
computer.}  This is not unreasonable for optical quantum computers,
since photons move around so easily, but is a poor assumption in
many other models, where qubits are constrained to interact only
with other qubits that are nearby in a 1-dimensional, 2-D, or
conceivably 3-D arrangement.
\item {\it Classical computation and measurement are fast and
reliable.}
In some systems, gate times are so fast that classical computation
cannot keep up; in others, measurement takes a long time compared to
the decoherence time, or is not possible on individual qubits.
\item {\it There is an ample supply of extra qubits.}  Perhaps someday
this will be true, but for the time being, extra qubits are at a
premium.
\item {\it Errors occur independently on separate gates and at separate
times, and $X$, $Y$, and $Z$ errors are equally likely, each with
probability $p/3$.}  This error model is known as the {\em
depolarizing channel}, and is a very common simple one to consider.
There are some variations on how it is extended to treat errors in
two-qubit gates, but generally there is a total probability $p$ of
error, which will likely affect both qubits involved in the gate. A
somewhat more realistic error model has errors occur with
probability $p$, independently on each gate, but does not specify
the probability of various different kinds of errors.  Even this
model will rarely hold to very high precision, as correlated errors
are likely between qubits which are located near each other.
\end{enumerate}
Much of the recent work on the threshold has focused on considering
different subsets of these assumptions or others like them, and
studying the effect on the threshold.  However, even if we make all
of the above assumptions, the threshold is still not a single
number, since different types of gates could have different error
rates.  Other work is devoted to studying fault-tolerance in
specific types of systems, such as ion traps or linear optical
quantum computation.

Another divide is between different methods of studying the
threshold.  Many of the results are derived using simulations to
estimate the threshold error rate.  If done correctly, simulations
can be very informative, but it is easy to make mistakes, for
instance to consider insufficiently large systems or insufficiently
general gate networks.  Since the simulations are usually performed
without a rigorous proof of correctness, there is always the
possibility that something critical has been omitted. The other main
approach is to mathematically prove a lower bound on the threshold
for a particular type of circuitry. This has the advantage that we
can be sure that a quantum computer that satisfies the appropriate
conditions will have a threshold at least that great.  Threshold
proofs can also frequently deal better with more general error
models.  However, proving a threshold generally requires making some
conservative simplifying assumptions that bring the resulting
threshold value lower than is achieved in a corresponding
simulation.  A factor of $10$ difference is common.  Most likely the
true threshold for a particular fault-tolerant scheme is somewhere
in between the simulated and proven threshold values.  There are
also a variety of analytical techniques which make some guesses
about which effects are most important and prove a threshold based
on those.  These techniques tend to produce a number somewhere
between the proof and the simulation.

The most optimistic estimates use {\em ancilla factories} to
effectively trade extra qubits for error tolerance.  Most of the
work is done on ancilla states, which are carefully checked and
discarded if found to be flawed.  Simulations suggest, using all of
the above assumptions, that the threshold can be pushed to at least
the range of $1\%$ -- $5\%$ using an extreme version of these
methods, depending on the relative error rates of different kinds of
gates. These schemes, however, cause a serious blow-up in the
overhead, to millions or billions of physical qubits for each
logical qubit, so should be regarded primarily as a theoretical
existence result for high thresholds.  Recent rigorous proofs of the
threshold using the extreme ancilla factory circuitry give a
threshold value of around $10^{-3}$, again an order of magnitude
less than the value suggested by simulations.

A fair amount of work has also now been done on systems with gates
constrained to act on nearest neighbor qubits in a 1-D or 2-D
lattice.  One-dimensional systems are difficult to deal with, but
there is still a fault-tolerant threshold.  In an almost 1-D system
(i.e., two parallel lines of qubits), the best current proof shows a
threshold of around $10^{-6}$ (in this case, I know of no published
superior simulation). In a 2-D system, the best simulation to date
gives a threshold of around $7 \times 10^{-3}$, while the best proof
gives around $2 \times 10^{-5}$, although using somewhat old
fault-tolerant techniques.  In general, then, working in a
two-dimensional system seems not to hurt the threshold too much,
perhaps by a factor of $2$ or $3$, but a one-dimensional system is
substantially inferior.

There has also been some study of the case where fast measurement or
reliable classical computation is unavailable. One solution is to
replace the classical computation by an equivalent quantum
computation. Since the quantum computer is unreliable, the
computation must be performed redundantly.  Luckily, it need only be
redundant in the sense of classical fault-tolerance, which can
tolerate a much higher error rate, so it appears that this approach
does not lower the threshold by very much.

We can also consider more general error models.  The depolarizing
channel is overly simplistic, and is unlikely to occur in any real
system.  A slightly more general model known as the {\em adversarial
probabilistic channel} is frequently considered in threshold proofs.
It assumes the locations (time and place) of errors are chosen
randomly and independently with probability $p$ per gate or time
step (with an erroneous two-qubit gate having errors on both
qubits), but that the actual type of error, $X$, $Y$, $Z$, or some
superposition, is chosen adversarially in such a way as to cause
maximum harm to the computation.  The adversary may even choose the
type of errors to be strongly correlated between error locations
(but the locations themselves are chosen independently).  Working
with the adversarial error model frequently results in a factor of
$2$--$3$ drop of the threshold compared to the depolarizing channel,
and is likely responsible for part of the difference between cited
numbers from proofs and from simulations.  (The adversarial error
model is not particularly amenable to simulation, whereas most
proofs have difficulty taking full advantage of the depolarizing
channel.)

However, even the adversarial error model does not capture the full
scope of realistic errors.  For instance, a slight over- or
under-rotation of a qubit leads to a coherent error, which cannot be
represented as a probability $p$ of some error (even an adversarial
one) and probability $1-p$ of no error.  In the most general case,
we should consider an environment interacting through a weak
Hamiltonian coupling to the system.  The environment may have an
indefinitely long memory and strong internal interactions.  Even in
such non-Markovian systems, there is threshold for fault-tolerance,
provided the system-bath Hamiltonian is bounded.  (It remains
unclear precisely what happens for unbounded systems once they are
regularized to avoid infinite energies.)

The provable threshold value for non-Markovian systems is much worse
than for probabilistic error models.  To compare properly, we must
convert the bound on the system-bath coupling which comes from the
proof to an error probability: namely, the probability that an ideal
state which undergoes the noisy gate will be projected by an ideal
measurement onto the correct output state.  Performing this
comparison, we find that the threshold for a non-Markovian system is
generally around the square of the threshold for the probabilistic
error model, with perhaps an additional factor of order $1$.

However, the threshold is unlikely to be that bad in reality.
Coherent errors, including non-Markovian ones, are dangerous in
principle because the amplitudes of errors can add up coherently,
causing the error probabilities to accumulate as the square of the
number of noisy gates rather than linearly.  In a real system, the
errors are unlikely to be able to arrange their phases so perfectly,
both because the environment is not truly an adversary, and because
the fault-tolerant protocol is constantly mixing things up.  If the
phases of consecutive errors are effectively randomized, the random
walk in phase causes probabilities to again accumulate linearly,
bringing the threshold back to roughly the level of probabilistic
errors.  It is my belief that this will be the case in practice,
although there is as yet no rigorous analysis to back up that
belief.

Other tradeoffs in properties relating to fault-tolerance still need
more study.  While ancilla factories use extra ancillas to allow
higher gate error rates, we also have some indication that it is
possible to go the other direction, and reduce overhead at a modest
cost in threshold.  To what degree is this possible?  There has been
almost no study of the effects of correlated errors on the threshold
error rate, but in many realistic systems, some correlations are
likely, at least between neighboring qubits.  And then there is the
question of trading off multiple factors simultaneously.  For
instance, to what degree can we create ancilla factories that
operate in two spatial dimensions?  Does fast {\em classical}
communication alter this tradeoff?  We are still a long way from
having a full understanding of this sort of question.

\section{For Further Information and Acknowledgements}

This has only been a very brief introduction to the accomplishments
and challenges of fault-tolerant quantum computation, but there are
a number of references which go into much more detail about quantum
error correction and fault tolerance.  For a brief introductions to
quantum error correction and fault tolerance, I suggest
\cite{QECCintro} and \cite{FTintro}, respectively.  Chapter 10 of
Nielsen and Chuang~\cite{NC} gives a more detailed overview of
quantum error correction and fault-tolerant circuitry.
\cite{AGP} provides a recent detailed
proof of the threshold theorem, along with a discussion of
non-Markovian errors. For the modern extreme version of ancilla
factories, see \cite{Knill} or the more detailed proofs
\cite{Reichardt} and \cite{AGPd2}.  For fault tolerance in 2
dimensions, see~\cite{FT2D} or \cite{FT2Dsim}, and for the
1-dimensional case, see~\cite{FT1D}. Finally, for tradeoffs between
overhead and error rate, see~\cite{Steane}.  I have not attempted
here to give full historical credit for development of the ideas,
but instead list, in most cases, the most recent, most general, or
otherwise best expression of each idea.  Again, consult the
references for more information about the historical development of
quantum error-correcting codes and fault-tolerant quantum
computation.

The author is supported by NSERC of Canada and by CIFAR.


\begin{thebibliography}{99}

\bibitem{QECCintro} D.~Gottesman, ``An Introduction to Quantum Error
Correction,'' in {\it Quantum Computation: A Grand Mathematical
Challenge for the Twenty-First Century and the Millennium}, ed.\
S.~J.~Lomonaco, Jr., pp.~221--235 (American Mathematical Society,
Providence, Rhode Island, 2002), quant-ph/0004072.
\bibitem{FTintro} J.~Preskill, ``Reliable Quantum Computers,''
Proc.\ Roy.\ Soc.\ Lond.\ A {\bf 454} (1998) 385--410,
quant-ph/9705031.
\bibitem{NC} M.~Nielsen and I.~Chuang, {\it Quantum Computation and
Quantum Information} (Cambridge University Press, 2000).
\bibitem{AGP} P.~Aliferis, D.~Gottesman, and J.~Preskill, ``Quantum
Accuracy Threshold for Concatenated Distance-3 Codes,'' Quant.\
Information and Computation {\bf 6}, No.~2, 97--165 (2006),
quant-ph/0504218.
\bibitem{Knill} E.~Knill, ``Quantum Computing with Realistically
Noisy Devices,'' Nature {\bf 434}, 39--44 (2005); E.~Knill,
``Fault-tolerant Postselected Quantum Computation: Schemes,''
quant-ph/0402171.
\bibitem{Reichardt} B. Reichardt, ``Error-Detection-Based Quantum
Fault Tolerance Against Discrete Pauli Noise,'' Berkeley Ph.D.\
thesis, 1996, quant-ph/0612004.
\bibitem{AGPd2} P.~Aliferis, D.~Gottesman, and J.~Preskill,
``Accuracy threshold for postselected quantum computation,''
quant-ph/0703264.
\bibitem{FT2D} K.~M.~Svore, D.~P.~DiVincenzo, and B.~M.~Terhal,
``Noise Threshold for a Fault-Tolerant Two-Dimensional Lattice
Architecture,'' quant-ph/0604090.
\bibitem{FT2Dsim} R.~Raussendorf, J.~Harrington, K.~Goyal,
``Topological fault-tolerance in cluster state quantum
computation,'' New J.~of Physics {\bf 9}, 199 (2007),
quant-ph/0703143.
\bibitem{FT1D} A.~M.~Stephens, A.~G.~Fowler, L.~C.~L.~Hollenberg,
``Universal fault tolerant quantum computation on bilinear nearest
neighbor arrays,'' quant-ph/0702201.
\bibitem{Steane} A.~M.~Steane, ``Overhead and noise threshold of
fault-tolerant quantum error correction,'' Phys.\ Rev.\ A {\bf 68},
042322 (2003) [19 pages], quant-ph/0207119.

\end{thebibliography}
\end{document}